\documentclass[prd,aps]{revtex4}
\usepackage{epsfig}
\usepackage{graphics}
\usepackage{latexsym}
\usepackage{amsmath}
\usepackage{amssymb}
\usepackage{rotating}
\usepackage{slashed}

\begin{document}

\title{The First Moments of Nucleon Generalized Parton Distributions}

\author{P. Wang$^{a,b,c}$}
\author{A. W. Thomas$^{d}$}

\affiliation{ $^a$Institute of High Energy Physics, Chinese Academy
of Science, Beijing 100049, P. R. China}

\affiliation{ $^b$Theoretical Physical Center for Science Facilities
(TPCSF), CAS, Beijing 100049, P. R. China}

\affiliation{ $^c$Jefferson Laboratory, 12000 Jefferson Ave.,
Newport News, VA 23606 USA}

\affiliation{$^d$CSSM, School of Chemistry and Physics, University
of Adelaide, Adelaide SA 5005, Australia}

\begin{abstract}
We extrapolate the first moments of the generalized parton
distributions using heavy baryon chiral perturbation theory. The
calculation is based on the one loop level with the finite range
regularization. The description of the lattice data is satisfactory and the
extrapolated moments at physical pion mass are consistent with the
results obtained with dimensional regularization, although the extrapolation
in the momentum transfer to $t=0$ does show sensitivity to form factor effects
which lie outside the realm of chiral perturbation theory. We discuss the
significance of the results in the light of modern experiments as well as QCD
inspired models.

\end{abstract}
%
%
\maketitle
\section{Introduction}

The study of hadron structure and, in particular, the spin structure
of the proton is one of the most exciting challenges facing
modern  nuclear and particle physics.
Through measurements of high energy scattering of electroweak probes
one can measure the parton distribution
functions (PDFs), which describe
the number densities of partons with momentum fraction, $x$, in the
nucleon. The study of the spin and flavor dependence of the PDFs has
provided a wealth of data which has proven critical to our understanding
of the structure of the nucleon~\cite{Thomas:2001kw}.

In recent years, the extension of this concept to so-called generalized
parton distributions (GPDs) has attracted enormous
interest~\cite{Ji0,Belitsky0}. The GPDs are related to the amplitude for deeply
virtual Compton scattering where the initial and final photons have
different momenta. It is widely anticipated that the GPDs should provide
even more information concerning the internal structure of the nucleon.
Of particular relevance to our present work is the link between the low
moments of the GPDs and the angular momentum carried by each quark flavor
within the proton~\cite{Ji:1995cu}.
This is now widely considered essential to a satisfactory
resolution of the famous proton spin crisis~\cite{Ashman:1989ig,Bass:2004xa},
because successful models suggest that a great deal of
the proton spin is carried as quark orbital
angular momentum~\cite{Myhrer:2007cf,Schreiber:1988uw,Thomas:2008ga}.

There have been many theoretical and experimental studies of the
GPDs. On the theoretical side, much of the work has centered on the
most effect ways to parameterize them~\cite{Guidal}. There has also
been quite a bit of work on phenomenological models,
such as the MIT and cloudy bag model~\cite{Ji1,Pasquini}, the
constituent quark model~\cite{Boffi,Scopetta}, the NJL model
\cite{Mineo}, the light-front model~\cite{Choi1,Choi2}, the color
glass condensate model~\cite{Goeke} and the Bethe-Salpeter approach
\cite{Tiburzi,Theussl}. On the experimental side, various
groups have focussed on different kinematic ranges. ZEUS and H1 measured the
DVCS cross section for the first time, with $x_B$ in the very low
range $10^{-4} < x_B < 0.02$~\cite{ZEUS,HI}. COMPASS focussed on a
little larger $x_B$ -- from $\simeq 0.01$ to $\simeq 0.1$
\cite{COMPASS}. The data from HERMES is in the range $0.03 < x_B <
0.35$~\cite{HERMES1,HERMES2}. The study of the high $x_B$ domain,
where valence quarks should dominate,
requires high luminosity and a high energy electron beam. With the
JLab 12 GeV upgrade, $x_B$ will range up to 0.7, while the current
JLab data is in the range $0.15 < x_B < 0.55$
\cite{CLAS1,CLAS2,CLAS3}.

The GPDs have a close relationship with the form factors. By
integrating the GPDs with different powers of the momentum fraction
$x$, the GPDs can be transformed into Mellin moments. There has been
important work on the moments and form factors from various lattice
QCD collaborations~\cite{LHPC,QCDSF,QCDSFUK}, as well as chiral
perturbation theory \cite{Dorati1,Dorati2,Diehl}. As for many other
observables computed in lattice QCD, current lattice simulations
have been concentrated at quite large pion mass. While chiral
perturbation theory is only expected to be convergent at quite low
pion mass~\cite{Young:2002ib,Leinweber:2005xz}, in order to relate
the simulations at large quark mass to experimental data, the
lattice data has been extrapolated using covariant and heavy baryon
chiral perturbation theory with dimensional regularization -- e.g.,
in Ref.~\cite{LHPC}.

Based on the observation that {\it all} hadron properties show a
slow, smooth variation with quark mass above $m_\pi \sim 0.4$GeV,
suggesting that chiral corrections (pion loops) are strongly
suppressed there~\cite{Thomas:2002sj}, an alternative regularization
method, namely finite-range-regularization (FRR) has also been used
to extrapolate the lattice data. It was first applied in the
extrapolation of nucleon mass and magnetic
moment~\cite{Leinweber:1998ej,Leinweber0,Leinweber1}. The remarkably
improved convergence properties of the FRR expansion mean that
lattice data at large pion mass can be described very well and the
obtained nucleon mass at physical pion mass is close to the
experimental value. Later, the FRR method was applied to extrapolate
the vector meson mass, nucleon magnetic moments, form factors,
charge radii and strange form
factors~\cite{Alton,Armour,Young,Wang1,Wang2,Wang3,Leinweber2,Leinweber3}.
Finally, we note that FRR has the unique advantage that it provides
a natural connection between physical results and quenched lattice
data~\cite{Alton,Armour,Wang2,Wang3}.

In this paper, we will focus on the low moments of the GPDs. The
LHPC lattice data will be used as input for the extrapolation.
The contribution from the disconnected quark loops has
not been included in the
lattice QCD simulation, which means that part of the sea quark contribution
has been omitted~\cite{LHPC}. While this contribution cancels in the isovector
moments it could be very important in the isoscalar moments. Extensive
investigation of other nucleon properties suggests that this omission
might be more important at the physical quark mass than at the relatively
heavy masses where the lattice calculations were
made~\cite{Thomas:2002sj}. If this were the case,
the chiral extrapolation, which {\it does} include contributions from
disconnected diagrams, may yield a reasonable representation of the physical
values, even for isoscalar quantities. This is indeed the case for the octet
magnetic moments~\cite{Leinweber2,Leinweber:2005bz}, for example.
On the other hand, we are unable to quantify
the error associated with this procedure for isoscalar quantities and in
future it would clearly be preferable to be able to work with lattice
simulations which include the disconnected contributions.

The paper is organized as follows. In section II, we briefly
introduce the chiral Lagrangian which is used for the calculation of
the first moments. The first isovector and isoscalar moments will be
derived in section III. Numerical results are presented in section
IV and finally section V presents a summary of the results with some
discussion.

\section{Chiral Lagrangian}

The generalized parton distribution functions, $H^q(x,\xi,t)$ and
$E^q(x,\xi,t)$, are defined as
\begin{eqnarray} \nonumber
\int \frac{d\lambda}{2\pi}~e^{i\lambda x}~\langle p^{'},s^{'}|\psi
\left( -\frac{\lambda n}{2} \right) \slashed {n} \psi
\left(\frac{\lambda n}{2}
\right)|p,s\rangle &=& H^q(x,\xi,t)\bar{u}(p^{'},s^{'})\slashed{n}u(p,s) \\
&+& E^q(x,\xi,t)\bar{u}(p^{'},s^{'})\frac{i\sigma^{\mu\nu}n_\mu Q_\nu}{2M_N} u(p,s),
\end{eqnarray}
where $Q=p^{'}-p, t=Q^2, \xi=-\frac12n\cdot Q$ with n the light-like
vector satisfying $n^2=0, \frac12 n\cdot Q=1$.
These GPDs can be transformed into Mellin moments/form factors
by integrating with different powers of the momentum fraction, $x$.
The zero-th order moments correspond to the Dirac and Pauli form
factors as
\begin{equation}
\int_{-1}^{1} dx x^0 H^q(x,\xi,t) = F_1^q(t),
\end{equation}
\begin{equation}
\int_{-1}^{1} dx x^0 E^q(x,\xi,t) = F_2^q(t).
\end{equation}
The Dirac and Pauli as well as the electric and magnetic form
factors have been discussed widely in the literature. In this paper,
we focus on the first moments of the nucleon GPDs:
\begin{equation}
\int_{-1}^1 dx x H^q(x,\xi,t) = A_{2,0}^q(t) + (-2\xi)^2
C_{2,0}^q(t),
\end{equation}
\begin{equation}
\int_{-1}^1 dx x E^q(x,\xi,t) = B_{2,0}^q(t) - (-2\xi)^2
C_{2,0}^q(t).
\end{equation}

The form factors $A$, $B$ and $C$ can be calculated with the
following equation:
\begin{equation}\label{current}
i\langle
p^{'}|\bar{q}\gamma_{\{\mu}\overleftrightarrow{D}_{\nu\}}q|p\rangle
= u(p^{'})\left[A_{2,0}^q(Q^2)\gamma_{\{\mu}\bar{p}_{\nu\}}
-\frac{B_{2,0}^q(Q^2)}{2M_N}Q^\alpha i
\sigma_{\alpha\{\mu}\bar{p}_{\nu\}} +
\frac{C_{2,0}^q(Q^2)}{M_N}Q_{\{\mu}Q_{\nu\}}\right]u(p),
\end{equation}
where the bracket, $\{...\}$, denotes the symmetrized and traceless
combination of all indices in the operator. $M_N$ is the nucleon
mass and $\bar{p}$ is the sum of the initial and final momenta. As one
can see from the above equation, the first moments of the GPDs can
be calculated by inserting into the nucleon states a tensor current
which interacts with the external tensor field.

It is convenient to define the isospin scalar and vector form
factors X (X=A, B or C) with the combination of each quark's
contribution:
\begin{equation}
X_{2,0}^{u+d}(Q^2) = X_{2,0}^u(Q^2) + X_{2,0}^d(Q^2),
\end{equation}
\begin{equation}
X_{2,0}^{u-d}(Q^2) = X_{2,0}^u(Q^2) - X_{2,0}^d(Q^2).
\end{equation}
In chiral perturbation theory, the interaction between tensor
current and the external tensor field as well as the baryon-meson
interaction can be written in a series of powers of momentum of the
tensor and meson fields ~\cite{Dorati1}. The lowest order Lagrangian
is
\begin{eqnarray}\nonumber
{\cal L}^{(0)} &=& \frac12 \bar{\psi}_N
\left\{i\left[\frac{a_{2,0}^v}{2}u^{\dag}V_{\mu\nu}^3\tau^3 u +
\frac{a_{2,0}^v}{2} u V_{\mu\nu}^3\tau^3 u^\dag +\frac{\Delta
a_{2,0}^v}{2}u^{\dag}V_{\mu\nu}^3\tau^3 u\gamma_5 \right . \right . \\
&-&\left . \left . \frac{\Delta a_{2,0}^v}{2} u V_{\mu\nu}^3\tau^3
u^\dag\gamma_5
+a_{2,0}^sV_{\mu\nu}^0\right]\gamma^{\{\mu}\overleftrightarrow{D}^{\nu\}}\right\}\psi_N,
\end{eqnarray}
where $V_{\mu\nu}^i$ and $V_{\mu\nu}^0$ are the isovector and
isoscalar tensor fields. The covariant derivative
$\overleftrightarrow{D}^\mu$ is defined as
$\overleftrightarrow{D}^\mu =\overrightarrow{D}^\mu -
\overleftarrow{D}^\mu$, where $\overrightarrow{D}^\mu = D^\mu
=\partial^\mu + \frac12\left[u^{\dag},
\partial^\mu u \right]$. $U=u^2$ is the non-linear realization of
the Goldstone boson field. As in Ref.~\cite{Dorati1}, the parity-odd
tensor interaction term is also included. In the above equation,
$\Delta a_{2,0}^v$ is only poorly known and it is related to the
spin-dependent analogue of the mean momentum fraction, namely
$<\Delta x>_{u-d}$.

The ${\cal O}(p^1)$ part of the interaction Lagrangian is expressed
as
\begin{eqnarray}\nonumber
{\cal L}^{(1)} &=& \bar{\psi}_N \left\{i\gamma^\mu D_\mu - M_N +
\frac{g_A}{2}\gamma^\mu\gamma_5u_\mu +\left(
\frac{ib_{2,0}^v}{8M_N}\left[D_\alpha,u^{\dag}V_{\mu\nu}^3\tau^3 u +
u V_{\mu\nu}^3\tau^3 u^\dag\right]\sigma^{\alpha\{\mu}D^{\nu\}}
+h.c.\right) \right . \\
&+& \left . \left(
\frac{ib_{2,0}^s}{4M_N}\left[\partial_\alpha,V_{\mu\nu}^0\right]\sigma^{\alpha\{\mu}D^{\nu\}}
+h.c.\right) \right\}\psi_N,
\end{eqnarray}
where $u_\mu$ is defined as $u_\mu=iu^\dag \partial_\mu u^\dag$.
The ${\cal O}(p^2)$ part of the interaction can be written as
\begin{eqnarray}\label{L2}\nonumber
{\cal L}^{(2)}&=&F_\pi^2Tr\left[\partial^{\{\mu} U^\dag
\partial^{\nu\}} U x_\pi^0
V_{\mu\nu}^0\right]-\frac{c_{2,0}^v}{2M_N}\bar{\psi}_N\left[D^{\{\mu},\left[D^{\nu\}},u^{\dag}V_{\mu\nu}^3\tau^3
u + u V_{\mu\nu}^3\tau^3 u^\dag\right]\right]\psi_N \\
&-&
\frac{c_{2,0}^s}{M_N}\bar{\psi}_N\left[\partial^{\{\mu},\left[\partial^{\nu\}},V_{\mu\nu}^0
\right]\right]\psi_N,
\end{eqnarray}
where the first part of this Lagrangian is the interaction between
the pion fields and the external isoscalar tensor field. $x_\pi^0$
is the momentum fraction of the pion carried by quarks. Its value is
less than 1 since some of the momentum of the pion is carried by
gluons. In the preceeding equations, $a_{2,0}^v$, $a_{2,0}^s$,
$b_{2,0}^v$, $b_{2,0}^s$, $c_{2,0}^v$ and $c_{2,0}^s$ are the low
energy constants which can be determined from the lattice data.

In our calculation, we also include the $\Delta$ intermediate
baryon, which is known to be crucial in the calculation of
spin dependent quantities.
The interaction between the $\Delta$ and the external tensor field can be
written as
\begin{eqnarray}\nonumber
{\cal L}_{\Delta}&=& -ia_{2,0}^{v,d}T^\alpha \gamma^{\{\mu}
\overleftrightarrow{D}^{\nu\}} T_\alpha V_{\mu\nu}^3 -
\frac{ib_{2,0}^{v,d}}{M_N}T^\alpha \overleftrightarrow{D}^\nu
T^\beta F_{\alpha\beta\nu}^3 +
\frac{b_{2,0}^{v,t}}{2M_N}(\bar{T}^\alpha \gamma^\beta
\overleftrightarrow{D}^\nu
 \psi+\bar{\psi}\gamma^\beta \overleftrightarrow{D}^\nu T^\alpha)F_{\alpha\beta\nu}^3 \\
&+& \frac{c_{2,0}^{v,d}}{M_N}T^\alpha
\left[D^{\{\mu},\left[D^{\nu\}},V_{\mu\nu}^3 \right]\right] T_\alpha
- ia_{2,0}^{s,d}T^\alpha \gamma^{\{\mu} \overleftrightarrow{\partial}^{\nu\}} T_\alpha V_{\mu\nu}^0
- \frac{ib_{2,0}^{s,d}}{M_N}T^\alpha \overleftrightarrow{\partial}^\nu T^\beta F_{\alpha\beta\nu}^0 \nonumber \\
&+& \frac{c_{2,0}^{s,d}}{M_N}T^\alpha
\left[\partial^{\{\mu},\left[\partial^{\nu\}},V_{\mu\nu}^0
\right]\right] T_\alpha,
\end{eqnarray}
where $F_{\alpha\beta\nu}^3 = [D_\alpha, V_{\beta\nu}^3]-[D_\beta,
V_{\alpha\nu}^3]$ and $F_{\alpha\beta\nu}^0 = [\partial_\alpha,
V_{\beta\nu}^0]-[\partial_\beta, V_{\alpha\nu}^0]$. The labels $d$
and $t$ in the low energy constants stand for decuplet and
transition, respectively. $T^\alpha$ are for decuplet fields which
have three flavor indices (they are not shown explicitly,
see for example, Ref.~\cite{James} for details), defined as
\begin{equation}
T_{111}=\Delta^{++}, ~~ T_{112}=\Delta^{+}, ~~ T_{122}=\Delta^{0},~~
T_{222}=\Delta^{-}.
\end{equation}

Within the framework of $SU(6)$ symmetry, there are relationships
between the octet and decuplet coefficients. For the isovector
coefficients:
\begin{eqnarray}\nonumber
a_{2,0}^{v,\Delta^{++}}=3a_{2,0}^v,~~~ a_{2,0}^{v,\Delta^{+}}=a_{2,0}^v,~~~,a_{2,0}^{v,\Delta^{0}}=-a_{2,0}^v,~~~,a_{2,0}^{v,\Delta^{-}}=-3a_{2,0}^v,
\\ \nonumber
b_{2,0}^{v,\Delta^{++}}=\frac{9}{5}b_{2,0}^v,~~~ b_{2,0}^{v,\Delta^{+}}=\frac{3}{5}b_{2,0}^v,~~~,b_{2,0}^{v,\Delta^{0}}=-\frac{3}{5}b_{2,0}^v,~~~,b_{2,0}^{v,\Delta^{-}}=-\frac{9}{5}b_{2,0}^v,\\
c_{2,0}^{v,\Delta^{++}}=3c_{2,0}^v,~~~ c_{2,0}^{v,\Delta^{+}}=c_{2,0}^v,~~~,c_{2,0}^{v,\Delta^{0}}=-c_{2,0}^v,~~~,c_{2,0}^{v,\Delta^{-}}=-3c_{2,0}^v.
\end{eqnarray}
For the isoscalar coefficients, we have:
\begin{eqnarray}\nonumber
a_{2,0}^{s,\Delta^{++}}=a_{2,0}^s,~~~ a_{2,0}^{s,\Delta^{+}}=a_{2,0}^s,~~~,a_{2,0}^{s,\Delta^{0}}=a_{2,0}^s,~~~,a_{2,0}^{s,\Delta^{-}}=a_{2,0}^s,
\\ \nonumber
b_{2,0}^{s,\Delta^{++}}=3b_{2,0}^s,~~~ b_{2,0}^{s,\Delta^{+}}=3b_{2,0}^s,~~~,b_{2,0}^{s,\Delta^{0}}=3b_{2,0}^s,~~~,b_{2,0}^{s,\Delta^{-}}=3b_{2,0}^s,\\
c_{2,0}^{s,\Delta^{++}}=c_{2,0}^s,~~~ c_{2,0}^{s,\Delta^{+}}=c_{2,0}^s,~~~,c_{2,0}^{s,\Delta^{0}}=c_{2,0}^s,~~~,c_{2,0}^{s,\Delta^{-}}=c_{2,0}^s.
\end{eqnarray}

With the Lagrangian, we can calculate the form factors. In the case
of the lowest moments, the electric form factor is related to the
contribution from the time component of the vector current, while
the magnetic form factor is related to the space component
contribution. For the first form factors, similar as in the case of
electric and magnetic form factors, one can also get three form
factors ${\cal A}_{2,0}$, ${\cal B}_{2,0}$ and ${\cal C}_{2,0}$ in
the heavy baryon formalism, whose relationships to the tensor
current are
\begin{equation}\label{J00}
J_{00}^q \equiv i\langle
p^{'}|\bar{q}\gamma_{\{0}\overleftrightarrow{D}_{0\}}|p \rangle =
\frac{3\bar{p}_0}2 {\cal A}_{2,0}^q(Q^2) + \frac{Q^2}{2M_N}{\cal
C}_{2,0}^q(Q^2),
\end{equation}
\begin{equation}\label{J33}
J_{33}^q \equiv i\langle
p^{'}|\bar{q}\gamma_{\{3}\overleftrightarrow{D}_{3\}}|p \rangle =
\frac{\bar{p}_0}2 {\cal A}_{2,0}^q(Q^2) + \frac{3Q^2}{2M_N}{\cal
C}_{2,0}^q(Q^2),
\end{equation}
\begin{equation}\label{J03}
J_{03}^q \equiv i\langle
p^{'}|\bar{q}\gamma_{\{0}\overleftrightarrow{D}_{3\}}|p \rangle =
\frac{i\bar{p}_0}{2M_N} {\cal B}_{2,0}^q(Q^2)(\vec{\sigma} \times
\vec{Q})_3.
\end{equation}

Within the heavy baryon formalism these three form factors are
related to the commonly used alternative form factors:
\begin{equation}\label{AA}
{\cal A}_{2,0}^q(Q^2) = A_{2,0}^q(Q^2) -
\frac{Q^2}{8M_N(E+M_N)}A_{2,0}^q(Q^2) -
\frac{Q^2}{4M_N^2}B_{2,0}^q(Q^2)
\end{equation}
\begin{equation}\label{BB}
{\cal B}_{2,0}^q(Q^2) = B_{2,0}^q(Q^2) + A_{2,0}^q(Q^2) -
\frac{Q^2}{8M_N(E+M_N)}B_{2,0}^q(Q^2)
\end{equation}
\begin{equation}\label{CC}
{\cal C}_{2,0}^q(Q^2) = C_{2,0}^q(Q^2) +
\frac{Q^2}{8M_N(E+M_N)}C_{2,0}^q(Q^2)
\end{equation}

\section{First moments of GPDs}

In this section, we will derive the formulas for the isovector and
isoscalar form factors ${\cal A}_{2,0}$, ${\cal B}_{2,0}$ and ${\cal
C}_{2,0}$. The one loop Feynman diagrams are shown in Fig.~1. The
solid lines are for the nucleons and $\Delta$. The dashed lines are
for the $\pi$ meson and the dotted lines are for the external tensor
current. Not all the diagrams have contribution to isoscalar or
isovector form factors. From Eq.~(\ref{L2}), one can see that there
is no corresponding interaction between the pion and the isovector tensor
field. This is because the $\tau$ matrix is traceless. For the
isoscalar form factors, the contributions of the $\pi^+$ and $\pi^-$ loops
in diagram c cancel. As a result, diagram c gives no
contribution to either the isoscalar or isovector form factors. In
Fig.~1, graphs b, d and f are of order ${\cal O}(m_\pi^2)$, while e is of
order ${\cal O}(m_\pi^3)$. The order of diagram a is dependent on
the moments. Its order is of ${\cal O}(m_\pi^3)$ for ${\cal
A}_{2,0}^s$ , ${\cal O}(m_\pi^2)$ for ${\cal B}_{2,0}^s$ and ${\cal
O}(m_\pi)$ for ${\cal C}_{2,0}^s$. We study the isovector form
factors first. For the isovector form factors, the diagram Fig.~1a
and 1c have no contribution and diagram 1e only contributes to
${\cal A}_{2,0}^v$.
\begin{center}
\begin{figure}[hbt]
\includegraphics[scale=0.5]{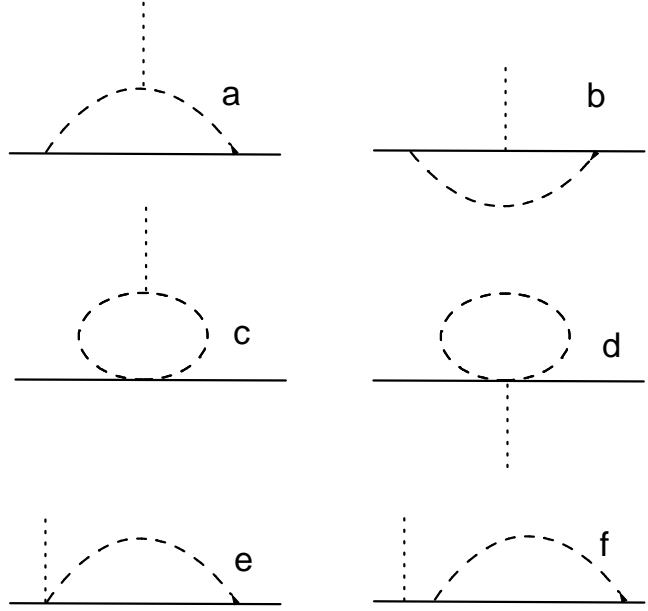}
\caption{One loop Feynman diagrams for the first moments of the GPDs.
The solid, dashed and dotted lines
are for nucleon(including $\Delta$), $\pi$ meson and external
tensor current, respectively.}
\end{figure}
\end{center}

The contribution of Fig.~1b, including wave function renormalisation
(Fig.~1f), is expressed as
\begin{equation}
{\cal A}_{2,0}^{v,b+f} = Za_{2,0}^v - \frac{g_A^2 a_{2,0}^v}{64
\pi^3 F_\pi^2} \int d^3 k \frac{\vec{k}^2
u^2(\vec{k})}{\omega^3(\vec{k})}- \frac{5{\cal C}^2 a_{2,0}^v}{72
\pi^3 F_\pi^2} \int d^3 k \frac{\vec{k}^2
u^2(\vec{k})}{\omega(\vec{k})(\omega(\vec{k})+\delta)^2},
\end{equation}
\begin{eqnarray}\nonumber
{\cal B}_{2,0}^{v,b+f} &=& Zb_{2,0}^v + \frac{g_A^2 b_{2,0}^v}{192
\pi^3 F_\pi^2} \int d^3 k \frac{\vec{k}^2
u^2(\vec{k})}{\omega^3(\vec{k})}+ \frac{5 {\cal C}^2 b_{2,0}^v}{216
\pi^3 F_\pi^2} \int d^3 k \frac{\vec{k}^2
u^2(\vec{k})}{\omega(\vec{k})(\omega(\vec{k})+\delta)^2} \\
&& \label{B20v} + \frac{g_A {\cal C} b_{2,0}^v}{90 \pi^3 F_\pi^2}
\int d^3 k \frac{\vec{k}^2
u^2(\vec{k})}{\omega(\vec{k})^2(\omega(\vec{k})+\delta)},
\end{eqnarray}
\begin{equation}
{\cal C}_{2,0}^{v,b+f} = Zc_{2,0}^v - \frac{g_A^2 c_{2,0}^v}{64
\pi^3 F_\pi^2} \int d^3 k \frac{\vec{k}^2
u^2(\vec{k})}{\omega^3(\vec{k})}- \frac{5{\cal C}^2 c_{2,0}^v}{72
\pi^3 F_\pi^2} \int d^3 k \frac{\vec{k}^2
u^2(\vec{k})}{\omega(\vec{k})(\omega(\vec{k})+\delta)^2},
\end{equation}
where $u(\vec{k})$ is the ultra-violet regulator and
$\omega({\vec{k}}) = \sqrt{\vec{k}^2+m_\pi^2}$. $\delta$ is the mass
difference between nucleon and $\Delta$. In the above equations,
${\cal C}$ is the nucleon-$\Delta$-$\pi$ coupling constant
\cite{Wang2}. The wave function renormalisation constant $Z$ is
expressed as
\begin{equation}
Z = 1 - \frac{3g_A^2}{64 \pi^3 F_\pi^2} \int d^3 k \frac{ \vec{k}^2
u^2(\vec{k})}{\omega^3(\vec{k})}- \frac{{\cal C}^2}{24 \pi^3
F_\pi^2} \int d^3 k \frac{\vec{k}^2
u^2(\vec{k})}{\omega(\vec{k})(\omega(\vec{k})+\delta)^2}.
\end{equation}
The first integrals in the above formulas are from the intermediate
nucleon contribution, while the second integrals are from the
intermediate decuplet contribution. The third integral in
Eq.~(\ref{B20v}) is from the $N-\Delta$ transition. This transition
contribution only exists for ${\cal B}_{2,0}^v$.

The contribution of Fig.~1d is obtained as
\begin{equation}
{\cal A}_{2,0}^{v,d} = - \frac{a_{2,0}^v}{16 \pi^3 F_\pi^2} \int d^3
k \frac{ u^2(\vec{k})}{\omega(\vec{k})},
\end{equation}
\begin{equation}
{\cal B}_{2,0}^{v,d} = - \frac{b_{2,0}^v}{16 \pi^3 F_\pi^2} \int d^3
k \frac{ u^2(\vec{k})}{\omega(\vec{k})},
\end{equation}
\begin{equation}
{\cal C}_{2,0}^{v,d} = - \frac{c_{2,0}^v}{16 \pi^3 F_\pi^2} \int d^3
k \frac{ u^2(\vec{k})}{\omega(\vec{k})},
\end{equation}
Fig.~1e only contributes to the form factor ${\cal A}_{2,0}$ expressed as
\begin{equation}
{\cal A}_{2,0}^{v,e} = \frac{g_A \Delta a_{2,0}^v}{48 \pi^3 F_\pi^2
M_N} \int d^3 k \frac{\vec{k}^2 u^2(\vec{k})}{\omega^2(\vec{k})}.
\end{equation}
{}From
the formulas one can see that in the heavy baryon formalism, the
lowest order loop contribution to the isovector form factors are
$Q^2$ independent.

The total expression for the form factors can be written as
\begin{equation}\label{vtotal}
{\cal G}_{2,0}^v(Q^2,m_\pi^2) = Z(g_{2,0}^v + g_\pi^v m_\pi^2 +(g_t^v + g_{\pi,t}^v m_\pi^2) Q^2)
+ \sum_{i=a}^f {\cal G}_{2,0}^{v,i},
\end{equation}
where ${\cal G}_{2,0}^v$ stands for ${\cal A}_{2,0}^v$, ${\cal
B}_{2,0}^v$ and ${\cal C}_{2,0}^v$. $g_{2,0}^v$, $g_\pi^v$, $g_t^v$
and $g_{\pi,t}^v$ are the corresponding low energy constants which
are determined by fitting the lattice data. In particular,
$g_{2,0}^v$ is identical to $a_{2,0}^v$, $b_{2,0}^v$ and
$c_{2,0}^v$, correspondingly. The other terms in the above equations
are from the tree level Lagrangian of high order. For example, for
${\cal A}_{2,0}^v$, the $m_\pi^2$ dependent term can be obtained
from the interaction $\bar{\psi}_N
\left\{\left[u^{\dag}V_{\mu\nu}^3\tau^3 u + u V_{\mu\nu}^3\tau^3
u^\dag \right]\left <\chi_+\right
>\gamma^{\{\mu}\overleftrightarrow{D}^{\nu\}}\right\}\psi_N$, where
$\chi_+ = u^{\dag}\chi u^{\dag}+u\chi^{\dag}u$ and $\chi=2B{\cal
M}$. $B$ is the chiral condensate and ${\cal M}$ is the quark mass
matrix. $\left < ... \right >$ denotes the trace in flavor space.
The $Q^2$ dependent term comes from the interaction $\bar{\psi}_N
\left\{\left[D^\alpha,\left[D_\alpha,u^{\dag}V_{\mu\nu}^3\tau^3 u +
u V_{\mu\nu}^3\tau^3 u^\dag
\right]\right]\gamma^{\{\mu}\overleftrightarrow{D}^{\nu\}}\right\}\psi_N$.

Since in the heavy baryon formalism, the loop contribution is $Q^2$
independent, the $Q^2$ dependence appears only in the analytic part.
One can also fit the data versus $Q^2$ at fixed pion mass. In that
case Eq.~(\ref{vtotal}) becomes:
\begin{equation}\label{vfixm}
{\cal G}_{2,0}^v(Q^2)|_{fixed ~ m_\pi^2} = Z(h_1^v +h_2^v Q^2) +
\sum_{i=a}^f {\cal G}_{2,0}^{v,i},
\end{equation}
where $h_1^v$ and $h_2^v$ are fitted independently for each pion mass.

For the isoscalar form factors, Fig.~1a, Fig.~1b and Fig.~1f
contribute, while fig.~1c and 1d give no contribution. The reason is
that in Fig.~1c and 1d, the $\pi^+$ and $\pi^-$ loops cancel each
other exactly. The contributions from 1b and 1f are
\begin{equation}
{\cal A}_{2,0}^s = Za_{2,0}^s + \frac{3g_A^2 a_{2,0}^s}{64 \pi^3
F_\pi^2} \int d^3 k \frac{\vec{k}^2
u^2(\vec{k})}{\omega^3(\vec{k})}+ \frac{{\cal C}^2 a_{2,0}^s}{8
\pi^3 F_\pi^2} \int d^3 k \frac{\vec{k}^2
u^2(\vec{k})}{\omega(\vec{k})(\omega(\vec{k})+\delta)^2},
\end{equation}
\begin{equation}
{\cal B}_{2,0}^s = Zb_{2,0}^s - \frac{3g_A^2 b_{2,0}^s}{192 \pi^3
F_\pi^2} \int d^3 k \frac{\vec{k}^2
u^2(\vec{k})}{\omega^3(\vec{k})}- \frac{5{\cal C}^2 b_{2,0}^s}{72
\pi^3 F_\pi^2} \int d^3 k \frac{\vec{k}^2
u^2(\vec{k})}{\omega(\vec{k})(\omega(\vec{k})+\delta)^2},
\end{equation}
\begin{equation}
{\cal C}_{2,0}^s = Zc_{2,0}^s + \frac{3g_A^2 c_{2,0}^s}{64 \pi^3
F_\pi^2} \int d^3 k \frac{\vec{k}^2
u^2(\vec{k})}{\omega^3(\vec{k})}+ \frac{{\cal C}^2 c_{2,0}^s}{8
\pi^3 F_\pi^2} \int d^3 k \frac{\vec{k}^2
u^2(\vec{k})}{\omega(\vec{k})(\omega(\vec{k})+\delta)^2}.
\end{equation}
Again, the first integrals of the above three formulas are from the
nucleon and the second integrals are from the $\Delta$ intermediate
state.

To evaluate Fig.~1a, we need to calculate the contribution of three components of
the tensor current defined in Eqs.~(\ref{J00})-(\ref{J03}). The expression for
these components are
\begin{equation}
J_{00}^s = - \frac{3g_A^2 x_\pi^0}{128 \pi^3 F_\pi^2} \int d^3 k
\frac{u(\vec{k})u(\vec{k}-\vec{q})[\vec{k}\cdot(\vec{k}-\vec{q})]^2
}{\omega^2(\vec{k})\omega^2(\vec{k}-\vec{q})}-\frac{{\cal C}^2
x_\pi^0}{48 \pi^3 F_\pi^2} \int d^3 k
u(\vec{k})u(\vec{k}-\vec{q})[\vec{k}\cdot(\vec{k}-\vec{q})]^2
f(\omega),
\end{equation}
\begin{eqnarray} \nonumber
J_{33}^s &=& - \frac{3g_A^2 x_\pi^0}{128 \pi^3 F_\pi^2} \int d^3 k
\frac{u(\vec{k})u(\vec{k}-\vec{q})[4k_z(k_z-q_z)-\vec{k}\cdot(\vec{k}-\vec{q})]\vec{k}\cdot(\vec{k}-\vec{q})}{\omega^2(\vec{k})\omega^2(\vec{k}-\vec{q})}
\\
&& -\frac{{\cal C}^2 x_\pi^0}{48 \pi^3 F_\pi^2} \int d^3 k
u(\vec{k})u(\vec{k}-\vec{q})[4k_z(k_z-q_z)-\vec{k}\cdot(\vec{k}-\vec{q})]\vec{k}\cdot(\vec{k}-\vec{q})
f(\omega),
\end{eqnarray}
\begin{eqnarray} \nonumber
J_{03}^s &=& \frac{i3g_A^2 x_\pi^0}{32 \pi^3 F_\pi^2} \int d^3 k
\frac{u(\vec{k})u(\vec{k}-\vec{q})k_z^2(\vec{\sigma}\times\vec{q})_z
}{\omega(\vec{k})\omega(\vec{k}-\vec{q})[\omega(\vec{k})+\omega(\vec{k}-\vec{q})]} \\
&& - \frac{i{\cal C}^2 x_\pi^0}{24 \pi^3 F_\pi^2} \int d^3 k
\frac{u(\vec{k})u(\vec{k}-\vec{q})k_z^2(\vec{\sigma}\times\vec{q})_z}{[\omega(\vec{k})+\delta][\omega(\vec{k}-\vec{q})+\delta][\omega(\vec{k})+\omega(\vec{k}-\vec{q})]},
\end{eqnarray}
where $f(\omega)$ is expressed as
\begin{equation}
f(\omega)=\frac{\omega(\vec{k})+\omega(\vec{k}-\vec{q})+\delta}{\omega(\vec{k})
[\omega(\vec{k})+\delta]\omega(\vec{k}-\vec{q})[\omega(\vec{k}-\vec{q})+\delta]
[\omega(\vec{k})+\omega(\vec{k}-\vec{q})]}.
\end{equation}
Both the intermediate nucleon and $\Delta$ states are included. One can see that this diagram
gives the $Q^2$ dependence of the form factors.
The total isoscalar form factors
${\cal A}_{2,0}^s$, ${\cal B}_{2,0}^s$ and ${\cal C}_{2,0}^s$
can be written in the same way as Eq.~(\ref{vtotal}).
The corresponding low energy constants
can also be determined by the lattice data.

\begin{center}
\begin{figure}[hbt]
\includegraphics[scale=0.5]{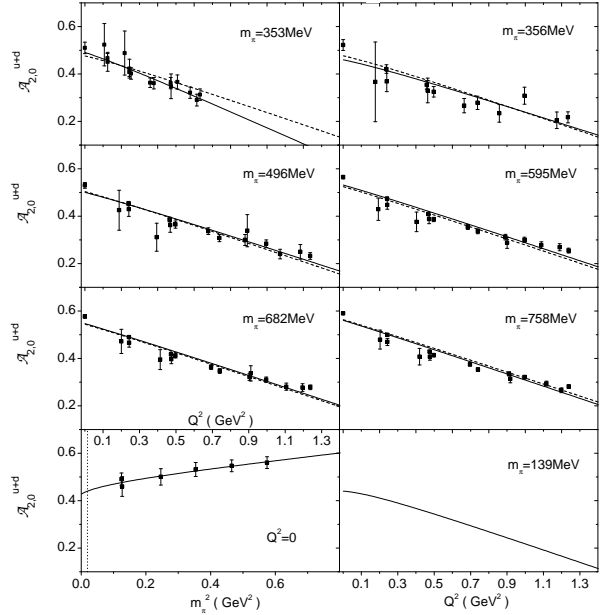}
\caption{The form factor ${\cal A}_{2,0}^{u+d}$ versus $Q^2$ at each
pion mass and versus pion mass at $Q^2=0$. The dashed and solid
lines correspond to the global and separate fit to the lattice data,
respectively.}
\end{figure}
\end{center}

\begin{center}
\begin{figure}[hbt]
\includegraphics[scale=0.5]{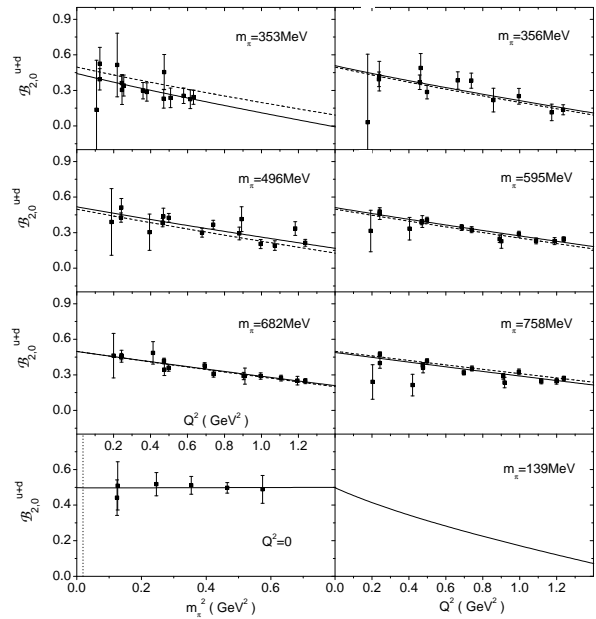}
\caption{The form factor ${\cal B}_{2,0}^{u+d}$ versus $Q^2$ at each
pion mass and versus pion mass at $Q^2=0$. The dashed and solid
lines correspond to the global and separate fits to the lattice
data, respectively.}
\end{figure}
\end{center}

\section{Numerical Results}

In the numerical calculations, the parameters are chosen to be $D=0.76$
and $F=0.50$ ($g_A=D+F=1.26$). The coupling constant ${\cal C}$ is
chosen to be $-2D$, as estimated by $SU(6)$ relations --- which
gives a similar value to that obtained from the hadronic decay width
of the $\Delta$.
Here the finite-range regulator is chosen to take the dipole form
\begin{equation}
u(\vec{k})=\frac1{(1+\vec{k}^2/\Lambda^2)^2},
\end{equation}
with $\Lambda = 0.8$ GeV. The regulator has been applied in our
previous work on nucleon mass, magnetic moments, form factors,
charge radii, etc. The other two parameters in the Lagrangian
$x_\pi^0$ and $\Delta a^v_{2,0}$ are chosen to be
0.7~\cite{Belitsky} and 0.21~\cite{Dorati1}, respectively.

All the lattice data had been transformed to a scale of $\mu^2=4$
GeV$^2$. We first study the isoscalar form factors. The form factor
${\cal A}_{2,0}^{u+d}$ is shown in Fig.~2. The lattice data are from
Ref.~\cite{LHPC}. The solid lines correspond to fitting the lattice
data with Eq.~(\ref{vfixm}) at each pion mass separately. The dashed
lines correspond to fitting the lattice data with Eq.~(\ref{vtotal})
for all the pion masses. We can see from the figure that the
difference between the solid and dashed lines is relatively small.
Both of them can extrapolate the lattice data very well. At each
pion mass, the lattice data shows a dependence on $Q^2$ which is
almost linear. In fact, the result of the Fig.~1 is $Q^2$
independent except for Fig.1a, which gives a little curvature to the
line. It is interesting that the momentum dependence of the first
form factor is not like the electromagnetic form factor which has a
dipole behavior. For the $m_\pi$ dependence, it is not like the
electromagnetic moment either. For example, For the magnetic moment,
Fig.~1a gives the leading order of ${\cal O}(m_\pi)$ contribution.
For ${\cal A}_{2,0}^{u+d}$, the leading order is of ${\cal
O}(m_\pi^2)$. This is because in this case, Fig.~1a are of ${\cal
O}(m_\pi^3)$. Fig.~1b and Fig.~1f are of ${\cal O}(m_\pi^2)$ which
is the same for the magnetic form factor. Therefore, the magnetic
form factor has a large curvature at small pion mass. At $Q^2=0$,
${\cal A}_{2,0}^{u+d}$ increases with increasing pion mass.

The form factor ${\cal B}_{2,0}^{u+d}$ is shown in Fig.~3. The solid
and dashed lines have the same meaning as in Fig.~2. Again, the lattice
data can be described very well. Similarly to ${\cal A}_{2,0}^{u+d}$,
${\cal B}_{2,0}^{u+d}$ has little curvature with increasing
$Q^2$. The form factor is not sensitive to the pion mass. For this
form factor, all the diagrams in Fig.~1 are of ${\cal O}(m_\pi^2)$
including Fig.~1a. At the physical pion mass, the value of ${\cal
B}_{2,0}^{u+d}$ at $Q^2=0$ is $0.497 \pm 0.089$.

\begin{center}
\begin{figure}[hbt]
\includegraphics[scale=0.5]{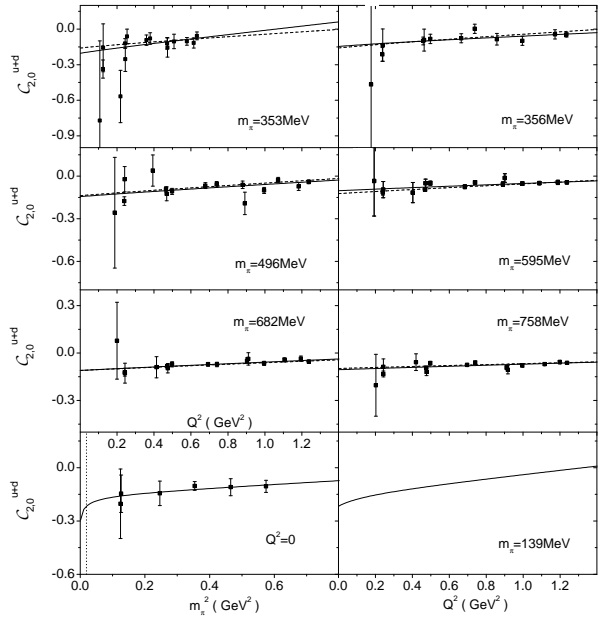}
\caption{The form factor ${\cal C}_{2,0}^{u+d}$ versus $Q^2$ at each
pion mass and versus pion mass at $Q^2=0$. The dashed and solid
lines correspond to the global and separate fits to the lattice
data, respectively.}
\end{figure}
\end{center}

\begin{center}
\begin{figure}[hbt]
\includegraphics[scale=0.5]{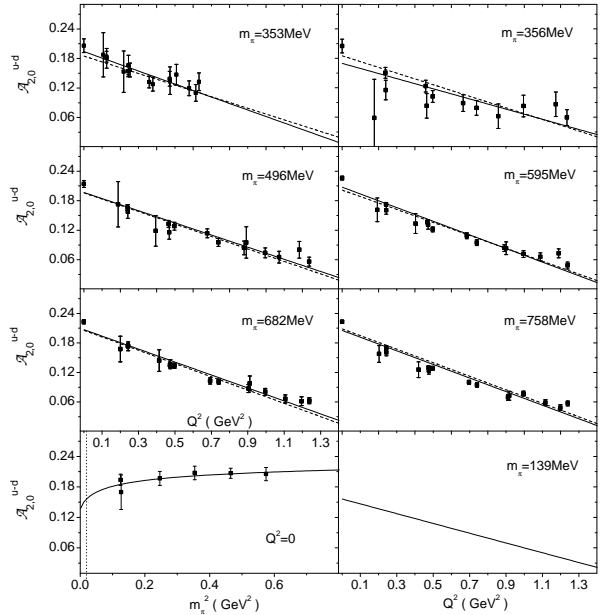}
\caption{The form factor ${\cal A}_{2,0}^{u-d}$ versus $Q^2$ at each
pion mass and versus pion mass at $Q^2=0$. The dashed and solid
lines correspond to the global and separate fits to the lattice
data, respectively.}
\end{figure}
\end{center}

{}From Ji's sum rule~\cite{Ji2}, the relationship between the
moments and the quark contribution to the total nucleon spin is
\begin{equation}
J_{u+d}=\frac12 [{A_{2,0}^s(Q^2=0)+B_{2,0}^s(Q^2=0)}] = \frac12
{\cal B}_{2,0}^s(Q^2=0).
\end{equation}
With the extrapolated value of ${\cal B}_{2,0}^s(Q^2=0)$
at the physical pion mass, we find $J_{u+d}=
0.249 \pm 0.045$. Since the total spin of the nucleon is $1/2$, it
is interesting that only $50\%$ of the nucleon spin is carried by
quarks. This is consistent with studies of the evolution of the total
quark angular momentum
from a scale typical of a valence-dominated quark
model~\cite{Thomas:2008ga}.

In Fig.~4  we show the form factor ${\cal C}_{2,0}^{u+d}$. Once
again the $Q^2$ dependence shows little curvature. However, this
time, the $m_\pi$ dependence has a visible curvature at the physical
pion mass and $Q^2=0$. This is because, unlike ${\cal
A}_{2,0}^{u+d}$ and ${\cal B}_{2,0}^{u+d}$, {}Fig.~1a has a leading
nonanalytic term of order ${\cal O}(m_\pi)$ for ${\cal
C}_{2,0}^{u+d}$. The absolute value of ${\cal C}_{2,0}^{u+d}$
decreases with increasing pion mass at $Q^2=0$.

Having discussed the isoscalar moments which describe the total
contribution of the $u$ and $d$ quark, we now turn to the isovector
part. The isovector moments describe the difference of the $u$ and
$d$ quark contributions. In this case, diagram a and c in Fig.~1
have no contribution and, as a result, the isovector form factors of
the loop are $Q^2$ independent. The form factor ${\cal
A}_{2,0}^{u-d}$ is shown in Fig.~5. Just as for the isoscalar form
factors, the diagrams of Fig.~1b, 1d and 1f are of ${\cal
O}(m_\pi^2)$. Because Fig.~1e is of ${\cal O}(m_\pi^3)$, the result
is not sensitive to the choice of $\Delta a_{2,0}^v$. For example,
at $Q^2=0$, the values of $A_{2,0}^{u-d}$ are 0.156 and 0.153 for
$\Delta a_{2,0}^v=0.21$ (phenomenological value) and 0.144 (fit
value), respectively \cite{Dorati1,Dorati2}. For the $Q^2$
dependence, from the figure one can see that the lines are well
described by a linear dependence on $Q^2$, which comes from the
choice of tree level contribution.

\begin{center}
\begin{figure}[hbt]
\includegraphics[scale=0.5]{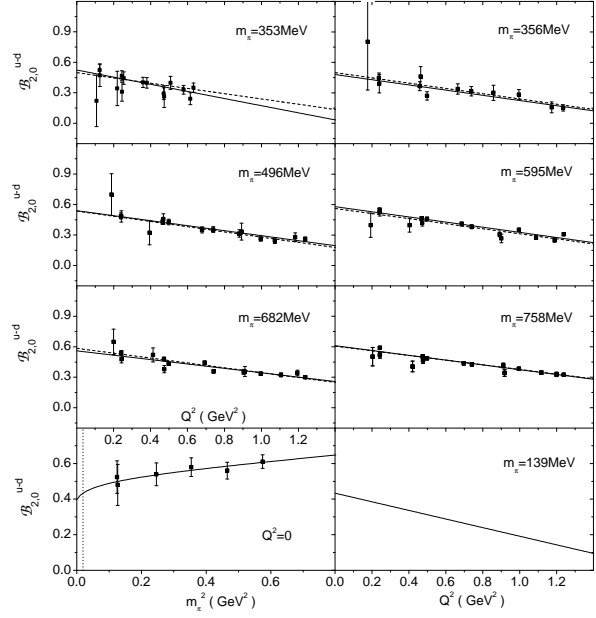}
\caption{The form factor ${\cal B}_{2,0}^{u-d}$ versus $Q^2$ at each
pion mass and versus pion mass at $Q^2=0$. The dashed and solid
lines correspond to the global and separate fits to the lattice
data, respectively.}
\end{figure}
\end{center}

\begin{center}
\begin{figure}[hbt]
\includegraphics[scale=0.5]{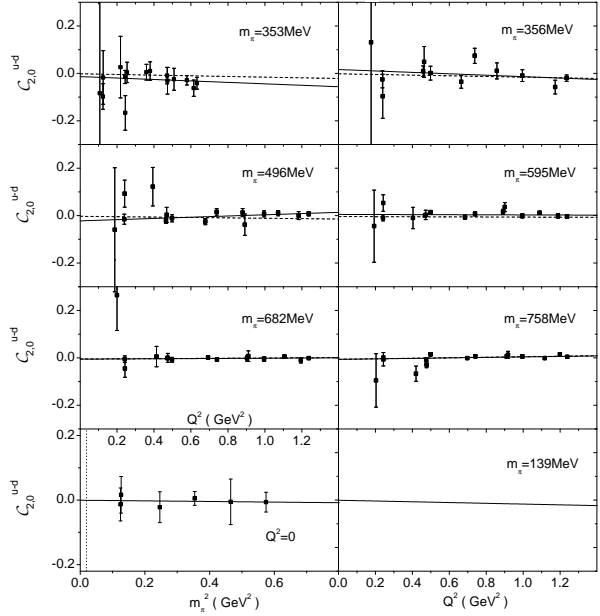}
\caption{The form factor ${\cal C}_{2,0}^{u-d}$ versus $Q^2$ at each
pion mass and versus pion mass at $Q^2=0$. The dashed and solid
lines correspond to the global and separate fits to the lattice
data, respectively.}
\end{figure}
\end{center}

In Fig.~6 we show the isovector form factor ${\cal B}_{2,0}^{u-d}$.
At $Q^2=0$, it increases with increasing $m_\pi^2$. At the physical
pion mass its value is about $0.433 \pm 0.071$. ${\cal
B}_{2,0}^{u-d}$ is close to ${\cal B}_{2,0}^{u+d}$, which means that
the $u$ quark is dominant for the proton spin while the $d$ quark
gives little contribution. The form factor ${\cal C}_{2,0}^v$ is
shown in Fig.~7. The value of ${\cal C}_{2,0}^v$ is around zero and
not sensitive to either $Q^2$ or $m_\pi^2$.

The values of ${\cal A}_{2,0}$, ${\cal B}_{2,0}$ and ${\cal
C}_{2,0}$ at the physical pion mass and $Q^2=0$ are shown in Table
I. Using Eqs.~(\ref{AA})-(\ref{CC}), one can extract the form
factors $A_{2,0}$, $B_{2,0}$ and $C_{2,0}$. In particular, at
$Q^2=0$, ${\cal A}_{2,0}=A_{2,0}$, ${\cal B}_{2,0}=A_{2,0}+B_{2,0}$
and ${\cal C}_{2,0}=C_{2,0}$. Compared with those in
Ref.~\cite{LHPC} extrapolated with the formulas in
Ref.~\cite{Dorati1}, our numerical results are close to their
results which used dimensional regularization. With the values for
the isoscalar and isovector moments in the Table, one can easily
find the $u$ and $d$ quark moments. For example, for the moment
${\cal A}_{2,0}$, the contribution from the $u$ quark is about as
twice large as that from the $d$ quark. This is because in the
proton, there are two $u$ quarks and one $d$ quark. The values of
the quark spin are $J_u=0.233 \pm 0.04$ and $J_d=0.016 \pm 0.04$ .
The $u$ quark dominance for the moment ${\cal B}_{2,0}$ can be
understood from the naive quark model for proton, where the $u$
quark spin is as four times large as the $d$ quark spin. For ${\cal
C}_{2,0}$, it seems that the $u$ and $d$ quarks yield almost the
same contribution.

\begin{table}
\caption{Low energy constants and moments at physical pion mass. The
results in the table are for the linear fit where the momentum
dependence of the tree level term is up to $Q^2$. For the dipole
fit, the first moments at $Q^2=0$ are about $10\% - 20\%$ larger.
The results of LHPC are also listed in the last column.}
\begin{center}
\begin{tabular}{||c|c|c|c|c|c|c||}
 & $g_{2,0}$ & $g_\pi$(GeV$^{-2}$) & $g_t$(GeV$^{-2}$) & $g_{\pi,t}$(GeV$^{-4}$) & ${\cal G}_{2,0}(0)$
& ${\cal G}_{2,0}^{\text {LHPC}}(0) [29]$
\\ \hline
$\mathcal{A}_{2,0}^{u+d}$ & 0.489 & 0.152 & $-0.266$ & 0.019 & $ 0.440 \pm 0.033$ & $ 0.520 \pm 0.024$ \\
\hline
$\mathcal{B}_{2,0}^{u+d}$ & 0.496 & 0.005 & $-0.339$ & 0.268 & $0.497 \pm 0.089$ & $ 0.425 \pm 0.086$ \\
\hline
$\mathcal{C}_{2,0}^{u+d}$ & $-0.152$ & 0.102 & 0.119 & $-0.159$ & $-0.217 \pm 0.103$ & $ -0.267 \pm 0.062$ \\
\hline $\mathcal{A}_{2,0}^{u-d}$ & 0.207 & 0.010 & $-0.139$ & $-0.009$ & $0.156 \pm 0.020$ & $ 0.157 \pm 0.010$ \\
\hline $\mathcal{B}_{2,0}^{u-d}$ & 0.526 & 0.158 & $-0.298$ & 0.111 & $0.433 \pm 0.071$ & $ 0.430 \pm 0.063$ \\
\hline
$\mathcal{C}_{2,0}^{u-d}$ & $-0.001$ & $-0.009$ & $-0.025$ & 0.062 & $ -0.001 \pm 0.050$ & $ -0.017 \pm 0.041$ \\
\end{tabular}
\end{center}
\end{table}

\begin{figure}[hbt]
\centering\includegraphics[width=0.5\columnwidth]{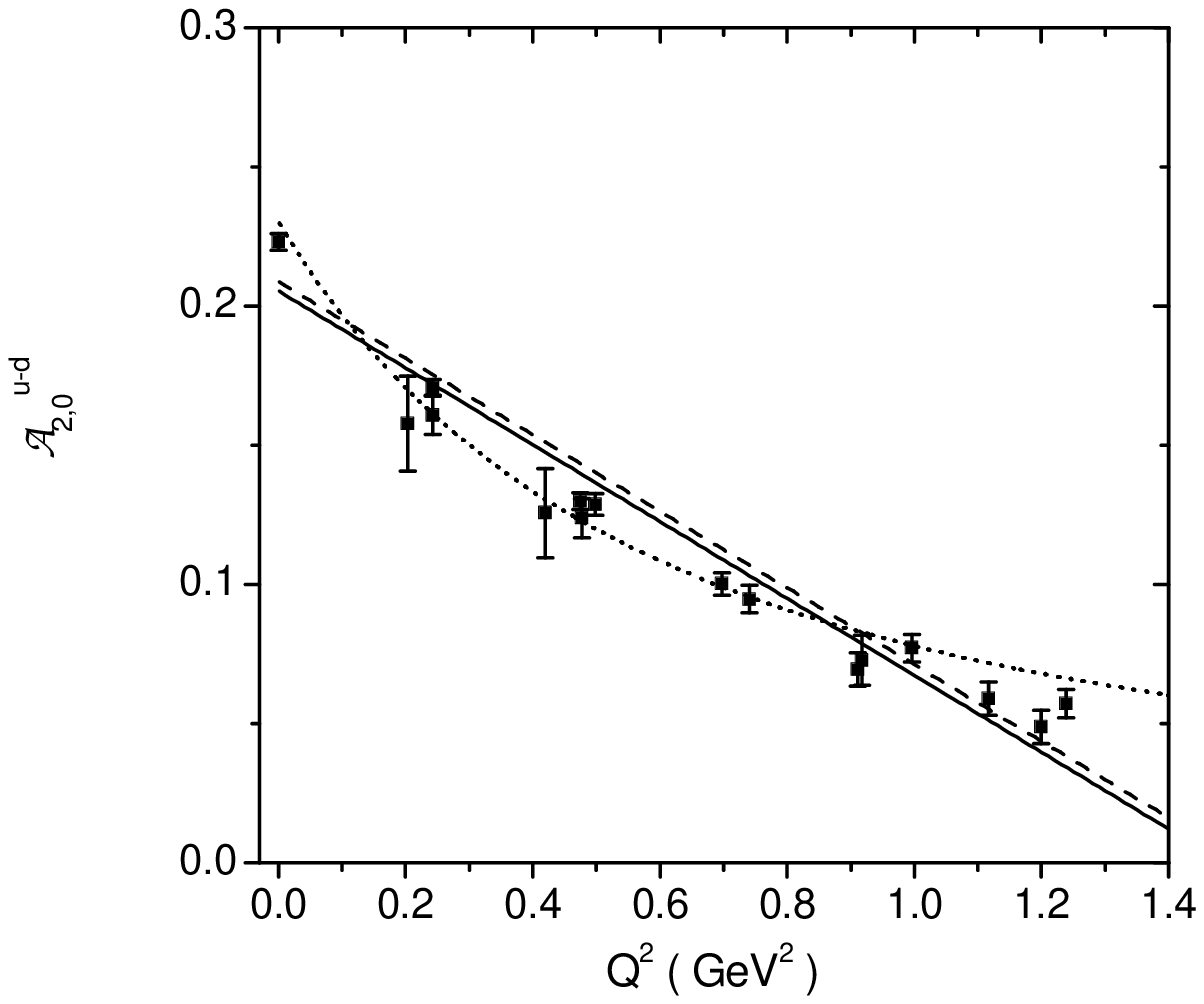}
\caption{The form factor ${\cal A}_{2,0}^{u-d}$ versus $Q^2$ at
$m_\pi=758$ MeV. The solid and dashed lines are for the linear fit
as in Fig.~5. The dotted line is for the dipole fit.}
\end{figure}

\begin{figure}[hbt]
\centering\includegraphics[width=0.5\columnwidth]{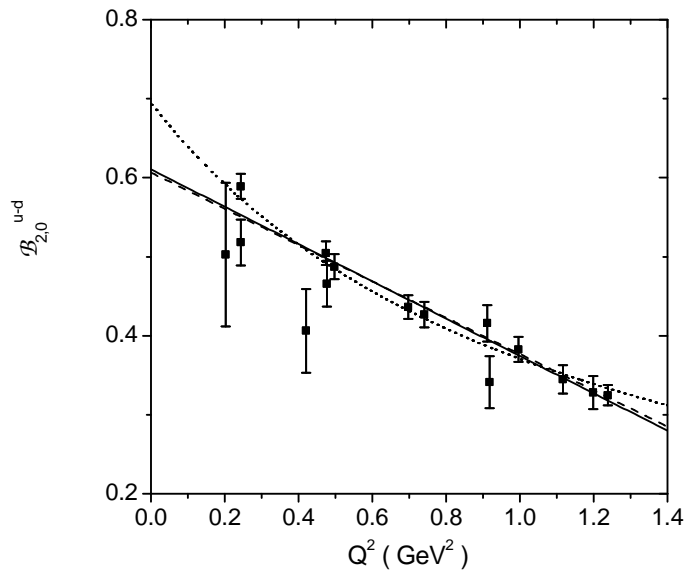}
\caption{The form factor ${\cal B}_{2,0}^{u-d}$ versus $Q^2$ at
$m_\pi=758$ MeV. The solid and dashed lines are for the linear fit
as in Fig.~6. The dotted line is for the dipole fit.}
\end{figure}

We should mention that our extrapolation results depend on the
choice of the tree-level contribution. In the previous fit, the
order of the momentum dependence of the tree-level contribution is up to
$Q^2$. This is supposed to be valid at low momentum transfer. We
know that for the electromagnetic form factors, the momentum
dependence has a dipole behavior with mass parameter around 0.71 GeV$^2$.
In the case of the axial form factor, which is perhaps more relevant for
the spin distribution, the phenomenological form factor is a 1 GeV dipole.
Since the actual lattice data extends over such a broad range of $Q^2$, a pure
linear dependence on $Q^2$ is difficult to justify.
Therefore, we have carried out another fit with a modified dipole form.
The $Q^2$ dependence of the first moments, i.e. Eq.~(\ref{vfixm}), is
now changed to the following expression:
\begin{equation}
{\cal G}_{2,0}^v(Q^2)|_{fixed ~ m_\pi^2} = \frac {Z(h_1^v +h_2^v
Q^2)}{(1 + Q^2/\Lambda^2)^2} + \sum_{i=a}^f {\cal G}_{2,0}^{v,i},
\end{equation}
where $\Lambda$ is chosen to be 1 GeV. While a pure dipole form would
be prefered phenomenologically, we note that the data from Ref.~\cite{LHPC}
tends to give much harder form factors for the axial channel than those
found in nature. The reason for this is not understood but our modified
form allows room to fit the lattice data while still including some
physically reasonable $Q^2$ dependence.

We show the momentum dependence of ${\cal A}_{2,0}^{u-d}$ and ${\cal
B}_{2,0}^{u-d}$ at $m_\pi=758$ MeV in Fig.~8 and Fig.~9. As a
comparison, the solid and dashed lines with the previous linear fit
are also shown. The dotted lines correspond to the fit with the
modified dipole form at tree level. The lattice data can be
reasonably described with both forms. For ${\cal B}_{2,0}^{u-d}$,
compared with the linear fit, the dipole fit gives a larger value at
$Q^2=0$. The situation is similar for the other pion masses. The
curve of mass dependence of ${\cal B}_{2,0}^{u-d}$ in Fig.~6 is
shifted up in the dipole fit. As a result, the moment at the
physical mass is increased from 0.433 to 0.53. For ${\cal
A}_{2,0}^{u-d}$, the difference between two fits is not as large as
for ${\cal B}_{2,0}^{u-d}$. This is because the lattice data for
${\cal A}_{2,0}^{u-d}$ at $Q^2=0$ impose a strong constraint in that
case. At the physical pion mass, its value changes from 0.156 to
0.17.

The dipole fit makes the absolute value of all the first moments
larger at $Q^2=0$. For ${\cal A}_{2,0}$, the difference is about
$10\%$ with the help of the lattice data at zero momentum. For
${\cal B}_{2,0}$ and ${\cal C}_{2,0}^{u+d}$, the difference is about
$20\%$. The value of ${\cal C}_{2,0}^{u-d}$ is still around zero.
The difference between these two separate fitting procedures
provides some indication of the systematic error in extracting this
important physical information from the lattice data.

\section{Summary}

Chiral perturbation theory with finite-range-regularization has been
applied to the problem of extrapolating lattice QCD data for GPD
moments to the physical pion mass and zero momentum transfer. For
the isovector form factors, the one loop contribution is of ${\cal
O}(m_\pi^2)$. For the isoscalar form factors, ${\cal A}^{u+d}_{2,0}$
and ${\cal B}^{u+d}_{2,0}$, the leading order is of ${\cal
O}(m_\pi^2)$, while for ${\cal C}^{u+d}_{2,0}$, the leading order is
of ${\cal O}(m_\pi)$. The lattice data were fitted both globally and
separately at each pion mass. At $Q^2=0$, the $m_\pi$ dependence of
the first moments (except ${\cal C}^{u+d}_{2,0}$) does not show a
big degree of curvature at small pion mass, which is quite different
from the zero-th moments (electromagnetic form factors). Overall the
level of agreement between the extrapolated results obtained using
dimensional regularization and FRR is satisfactory.

The $Q^2$ dependence, especially of ${\cal B}_{2,0}$, is mainly
determined by the tree level $Q^2$ behavior. In our first
calculation, we retained only the tree level terms  up to $Q^2$,
that is a linear dependence. On the other hand, the data has been
determined over a wide range of $Q^2$, up to 1.2 GeV$^2$.
Phenomenlogically one knows that the physical form factors exhibit a
considerable variation with $Q^2$ over such a range -- usually
described by a dipole form. In order to test the sensitivity to this
problem, we have re-done the fits to the lattice data using a
modified dipole form. In this case the first moments at $Q^2=0$ are
about $10\% - 20\%$ larger than those in the linear fit. We regard
this as a measure of one of the main systematic errors associated
with extracting information about the quark angular momentum.

In our calculation, the $\Delta$ contributions have been included
explicitly, because of its potential importance for spin
dependent quantities.
For the nucleon spin, the inclusion of the $\Delta$ changes the result
by $10\%$~\cite{Myhrer:2007cf}. However, for the extrapolation of the first
moments of GPDs, if the $\Delta$ is not included the low energy
constants are different. Since they are adjusted by fitting the lattice
data, the extrapolated moments at the physical pion mass
change by less than about $5\%$ when the $\Delta$ is included.

Using the extrapolated moment ${\cal B}_{2,0}$, we can extract
information concerning the quark contribution to the nucleon spin.
Our results are consistent with the current JLab and HERMES
experiments \cite{Ye,Mazouz}. On the other hand, the experimental
errors there are quite large at the present time. The comparison
with model predictions, such as those of
Refs.~\cite{Myhrer:2007cf,Thomas:2008ga}, is also quite
satisfactory, especially given the systematic errors associated with
the $Q^2$ dependence of ${\cal B}_{2,0}$ (discussed above) and the
dependence of the proton axial charge on lattice volume. As for the
other two moments, the $u$ quark contribution is twice as large as
that for the $d$ quark in ${\cal A}_{2,0}$, while they give similar
contributions to ${\cal C}_{2,0}$.
In comparing with data, it is important to note
the possible sensitivity of isoscalar quantities
to disconnected graphs which were not included in the LHPC simulation.
In the next few years we can look
forward not only to more accurate data on GPDs but lattice QCD data
on larger volumes and at lower quark masses and, ideally, including
disconnected terms. All of this offers
enormous promise towards unravelling the proton spin problem.

\section*{Acknowledgements}
This work was supported by DOE contract No. DE-AC06-05OR23177 and by the
Australian Research Council grant of an Australian
Laureate Fellowship to AWT.

\end{document}